\documentclass[aps,pre,twocolumn]{revtex4}

\usepackage{graphicx}
\usepackage{amsmath}

\begin{document}

\title{From Classical to Quantum Glasses with Ultracold Polar Molecules}
\author{Wolfgang Lechner}
\email{w.lechner@uibk.ac.at}
\author{Peter Zoller}
\affiliation{Institute for Quantum Optics and Quantum Information, Austrian Academy of
Sciences, 6020 Innsbruck, Austria}
\affiliation{Institute for Theoretical Physics, University of Innsbruck, 6020 Innsbruck,
Austria}
\date{\today }

\begin{abstract}
We study the dynamics of a bilayer system of ultracold polar molecules, which exhibits
classical and quantum glassy behavior, characterized by long tails in the
relaxation time and dynamical heterogeneity. In the proposed setup, quantum
fluctuations are of the order of thermal fluctuations and the degree of
frustration can be tuned by the interlayer distance. We discuss the possible observation of a glassy anomalous diffusion and dynamical heterogeneity in experiment using internal degrees of freedom of the molecules in combination with optical detection. 
\end{abstract}

\maketitle

The recent experimental realization of cold ensembles of polar molecules \cite{REVIEWCEHM} has
opened a new pathway to explore the dynamics of quantum many body systems
with strong, long-range and anisotropic polar interactions\cite{CARR,KKNI,OSPELKAUS,Yan,WEIDEMUELLER,HCN,Baranov,Lewenstein}. In
combination with low dimensional trapping geometries, this allows the
realization of stable many body phases, for example in the form of 1D wires
or 2D pancakes \cite{Buechler,Astra,Cooper,LEWENSTEINBOOK}, or as stacked pancakes
representing coupled multilayer systems \cite{Z_PRL_SELFASSEMBLY,BILAYERLUKIN,BILAYER,Potter,Knap}. Most of the
experimental and theoretical studies of polar molecular gases have focused
so far on the quantum degenerate regime, and on equilibrium phenomena,
including superfluid and crystal phases, quantum magnetism, and topological
phases (for a review see \cite{Baranov,Lewenstein}). Instead, our interest
below will be on non-equilibrium many-body dynamics. We will show that a
bilayer setup of ultracold polar bosonic molecules can feature a \emph{glass
phase}, and we present methods to prepare this phase and measure the
relevant order parameters with tools available with present experimental
setups. The unique feature and the theoretical challenge of glass physics
with cold molecular ensembles is the possibility to study the crossover from
classical to quantum glasses. 

\begin{figure}[ht]
\centerline{\includegraphics[width=7.5cm]{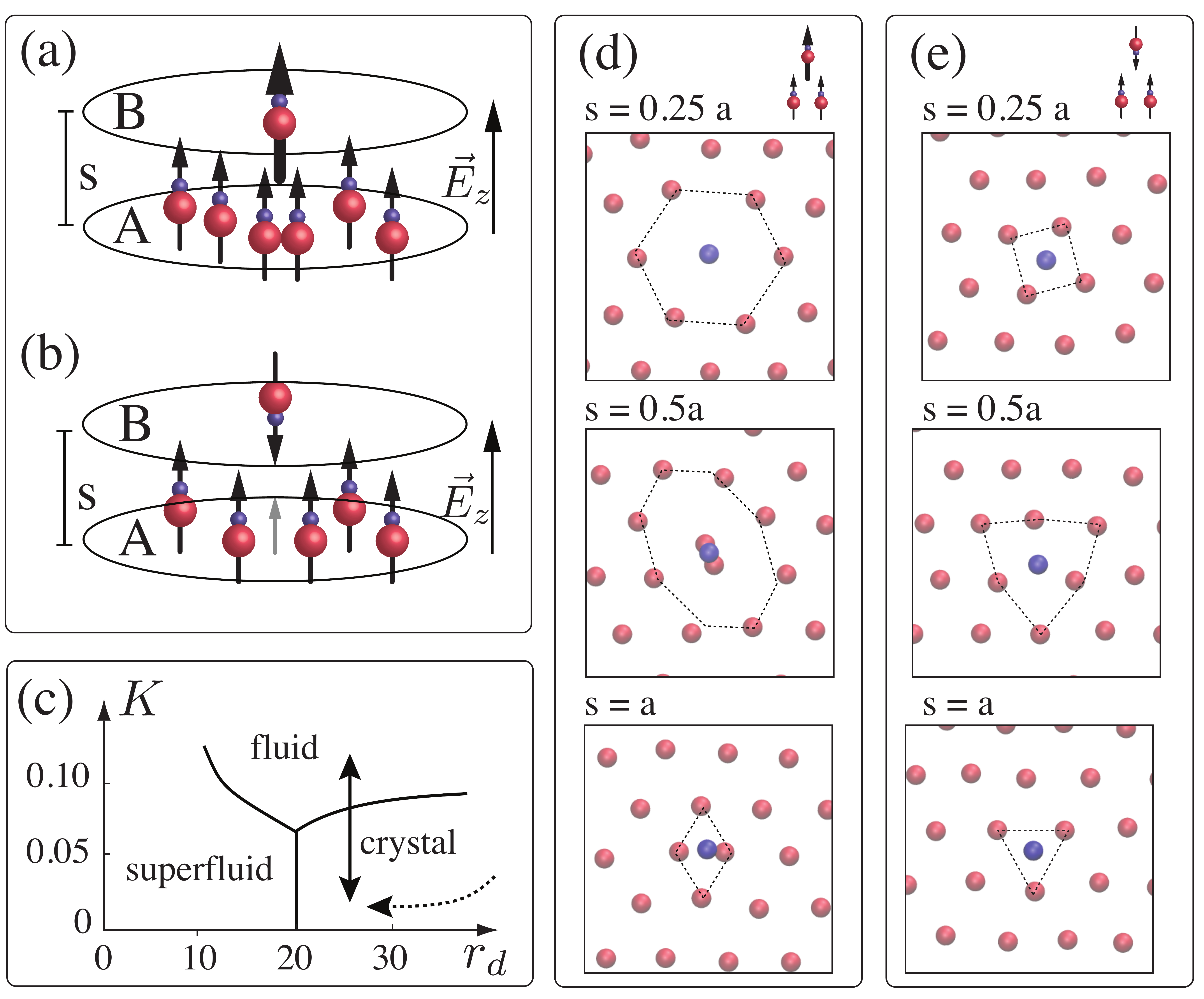}}
\caption{(Color online)  Bilayer setup of trapped polar
molecules with parallel dipole moments (a) and anti-parallel dipole moments (b). The lower layer represents a high-density dipolar crystal, the upper layer is loaded with low density (see text). A 
typical route to enter the glass phase is to quench the system from the liquid to a supercooled phase. In contrast, here, the glass phase is reached by decreasing the layer separation $s$. (c) 
Sketch of the phase diagram of a single layer of
bosonic polar molecules with the ratio between kinetic energy and interaction $K$ and $r_d$ from Ref. \cite{Buechler}. Note, that the limit $m \rightarrow \infty$ corresponds to the classical limit as the extension of the wavefunction vanishes with $\lambda_{dB} = h/(2 \pi m k_B T)^{1/2}$. The arrow indicates the parameter region of interest. (d) For parallel 
dipoles [panel (a)], the effective volume and symmetry of the defects (dimers,
trimers) depends on the interlayer separation. The ground state from a
classical simulated annealing calculation at $T=0$ results in patterns
include triangular, cubic and asymmetric structure where particles in the
crystal layer (red) are displaced by the molecules in the defect layer
(blue). (e) For anti-parallel dipoles [panel (b)] defect patterns can range
from triangular, cubic to fivefold symmetries.}
\label{fig:illustration}
\end{figure}

A glass phase is characterized by a plateau in the relaxation time-scale,
known as \textit{aging}, with exponentially increasing tails that prevent
the system from reaching its equilibrium state \cite{BINDERBOOK, LEWENSTEINBOOK, DEBENEDETTI2001,FRUSTRATION}. In a structural glass, global reorganization to 
the equilibrium is prevented from geometric frustration as a result of the dynamics \cite{Disorder}.
While large reorientation is slow, on a local scale, relaxation can be fast, a 
phenomenon known as \textit{dynamical heterogeneity} 
\cite{CHANDLER}.  In a classical glass this relaxation dynamics is dominated by thermal fluctuations, and a considerable understanding of
the glass transition has been gained from various theoretical methods \cite{BINDERBOOK,TANAKACRITICAL,KINETICISING,KINETIC,REPLICA,WOLYNES} and
experimental model systems \cite{REVWEEKS,WEEKSGLASS,KEIM2008}. In contrast,
the question of the influence of quantum fluctuations on the glass
relaxation dynamics is far less well understood. Recent theoretical studies indicate,
based on analytical \cite{NOVIKOV2013,RABANIMCT, WOLYNES} and numerical \cite{MARKLAND2011,MARKLAND2012,QUANTUMJAMMING} methods, that quantum
fluctuations can enhance but also inhibit the glass transition.

In this Letter, we propose and analyze a protocol to prepare and measure a
glass phase in a bilayer setup of cold polar molecules (see Figs.~\ref{fig:illustration}a and b)
in the regime of the cross over from a classical to a
quantum glass. We assume ultracold molecules prepared in their
electronic and rovibrational ground states, where a  static electric field 
$\mathbf{E\equiv }E\mathbf{e}_{z}$ oriented perpendicular to the trapping
layers in the $xy$-plane induces an electric dipole moment $\mathbf{d}\equiv d\mathbf{e}_{z}$. Thus the molecules will interact according strong, long-range and anisotropic dipole-dipole interaction. The stacked pancake potential of Figs.~
\ref{fig:illustration}a and b can be realized with a 1D optical lattice with layer separation $s$ controllable by laser parameters. For strong confinement the motion of the molecule within each layer 
can be described as an effective 2D dynamics, while the tunneling between adjacent layers will be suppressed by a sufficiently high
barrier. The Hamiltonian for our bilayer system thus has the form $H=H_{A}+H_{B}+H_{AB}$. Here $H_{A}$ ($H_{B}$) \ is the Hamiltonian for the
intralayer dynamics, 
\begin{equation}
H_{X}=\sum_{i\in X}\left( \frac{\mathbf{p}_{i}^{2}}{2m}+V_{T}(\mathbf{r}
_{i})\right) +\sum_{i,j\in X}\frac{d_{X}^{2}}{4\pi \epsilon _{0}}\frac{1}{r
_{ij}^{3}} \quad (X=A,B)  \label{eq:X}
\end{equation}%
as sum of kinetic energy and a tranverse trapping potential, and the (purely
repulsive) dipolar interaction with in-plane distance $r$.   The
interlayer interactions are
\begin{equation}
H_{AB}=\sum_{i\in A,j\in B}\frac{d_{A}d_{B}}{4\pi \epsilon _{0}}\frac{1}{R_{ij}^3} \left( 1 - \frac{3 s^{2}}{R_{ij}^{2}} \right).  \label{eq:AB}
\end{equation}%
with $s$ the layer separation and $R$ the distance between the molecules with $R^2 = r^2 + s^2$.
Note, that our model allows for different dipole moments $d_{A}$ and $d_{B}$ in layers $A$ and $B$, respectively. Different
effective dipole moments in the two layers can be achieved e.g.~by a
gradient in the electric field $E_z$ or by engineering of the interactions
using internal rotational degrees of freedom \cite{Baranov}. As we will see below, this feature is essential to observe clear signatures of a glass phase. 

\begin{figure}[ht]
\centerline{\includegraphics[width=8.5cm]{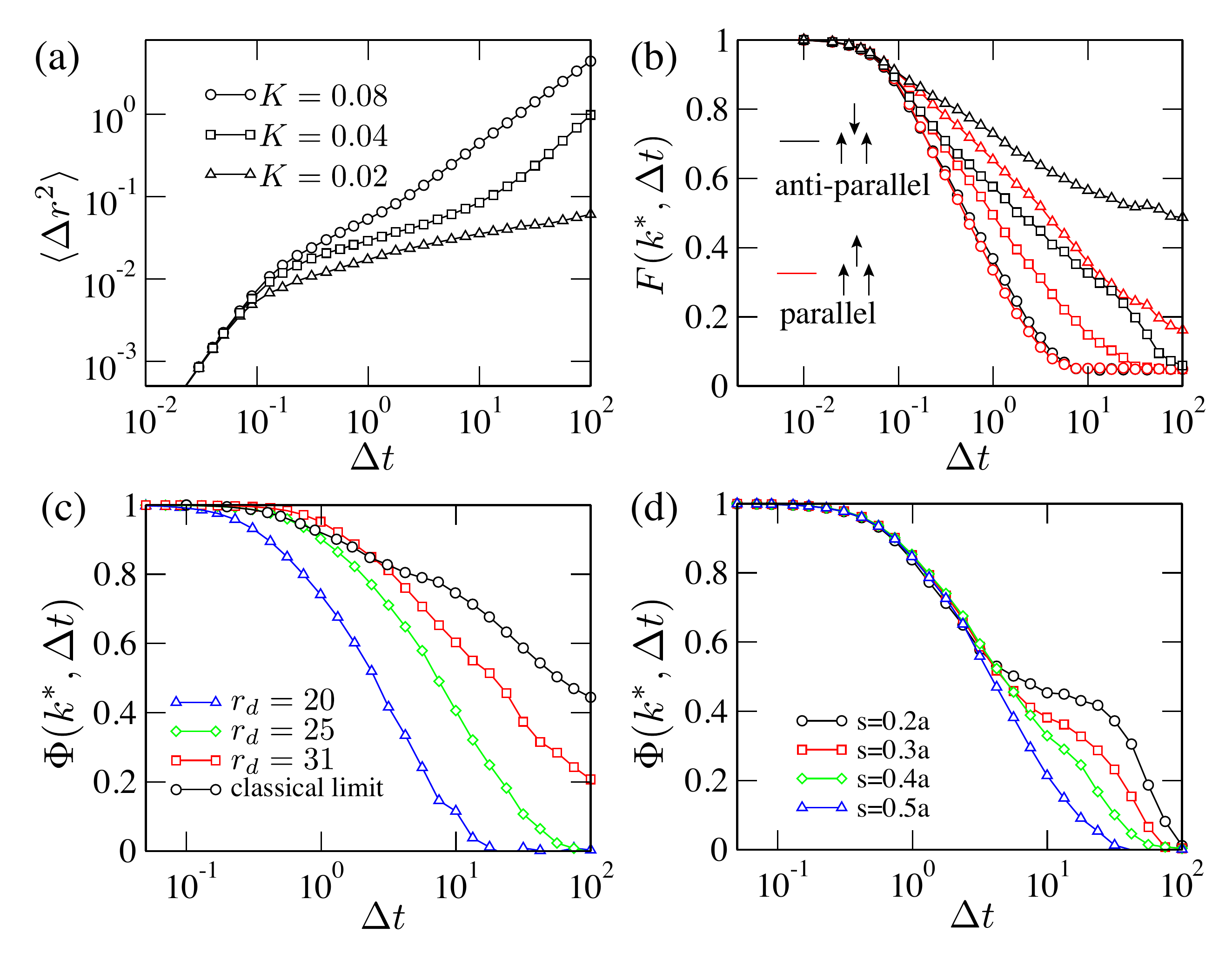}}
\caption{(Color online) Glass order parameters in the classical (top) and
semi-classical regime from path integral calculations (bottom). (a) In the
glass phase, the average mean squared displacement of the particles deviates
from the constant diffusion. In the anti-parallel setup the plateau extends over the whole sampling time
below $K<0.02$ of the combined system. (b) The relaxation of the time dependent structure factor $F(k^*,t)$ for $s=0.5a$ with $d_B=4 d_A$ (black) and $d_B=-d_A$ (red)
diverges when approaching the glass phase [symbols as in (a)]. (c)
Relaxation dynamics with initial temperature $K=0.01$ in the anti-parallel setup for various
choices of effective $r_d$ of the combined system. Parameters are chosen such that the classical limit $m
\rightarrow \infty$ enters a glass phase and for smaller $r_d$ the system melts
due to large quantum fluctuations (blue). This allows one to study the cross-over from a classical glass to a quantum phase. (d) Relaxation dynamics as a
function of the layer separation with effective $r_d \approx 20-25$ and $K=0.02$ in the
anti-parallel setup. While for distance $s = 0.5a$ the system is a liquid due to large quantum fluctuations
(blue) one reaches the glass phase when approaching $s=0.2a$ (black). The
number of particles used was $N_A=780$ and $N_B=200$ in the classical
simulations and $N_A=168$ and $N_B=50$ in the path integral simulations using $%
50$ time-slices per particle and periodic boundary conditions. Averages are taken
from $100$ independent runs. For LiCs a
time of $\Delta t = 10^2$ in reduced units corresponds to $\Delta t=3.4$
milliseconds.}
\label{fig:resultspanel}
\end{figure}

\begin{figure}[ht]
\centerline{\includegraphics[width=8cm]{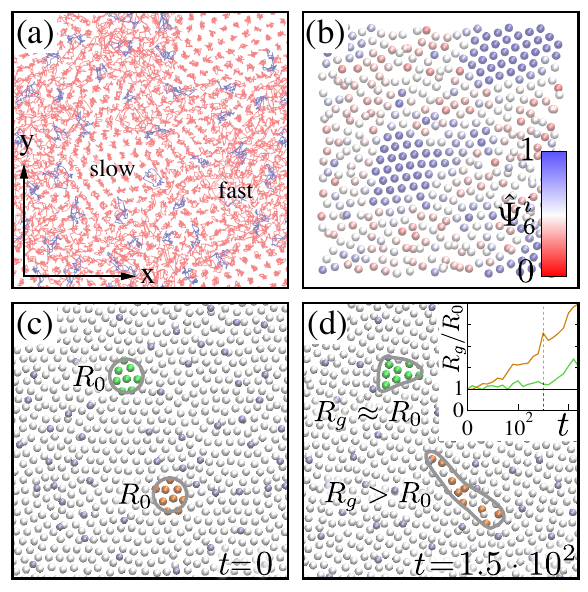}}
\caption{(Color online) Dynamical heterogeneity of a quantum glass in an AMO 
experiment: Molecules in an excited internal 
state act as \textit{marker molecules} to identify glassy dynamics. (a) Typical initial 
configuration of layer $A$ (red)  and layer $B$ (blue) depicted as the projection
of the paths onto real space. (b) The parameter $\psi_6^i$ measures the relative order in the vicinity of particle $i$ (for definition and details see SM \cite{supmat}). Here, regions of large order ($\psi_6^i \approx 1$) are regions of small classical diffusion and a small extension of the particle wavefunction. Regions of low order ($\psi_6^i \approx 0$) feature large diffusion and larger quantum fluctuations. Note, that dynamical heterogeneity is a more general feature of glasses, which is also present in systems without apparent partial crystallinity (see SM \cite{supmat}). 
(c) Snapshots of two possible initial choices of marker molecule positions (green and orange) at $\Delta t=0$ 
with $R_g=R_0$. The particle positions indicated by spheres
represent the maximum of the path probability in layer $A$ (gray) and layer $B$ (lightblue).
(d) After evolving the system in time, the radius of gyration of marker molecules after 
time $\Delta t=1.5 \times 10^2$ ($5 \text{ms}$) differs depending on the initial position.  
In the glass phase, due to dynamical heterogeneity,
the growth of $R_g$ strongly depends on the choice of the initial position. The cloud of 
maker molecules placed in an inactive region (green) of the glass follows the
dynamics of an amorphous crystal $R_g = R_0$ while molecules in a 
mobile region (orange) show a fast increase in size (inset).}
\label{fig:marker}
\end{figure}

The basic steps of our protocol to study glassy dynamics are as follows. (i)  We start with two uncoupled layers ($s$ large), where layer $A$ is in a high density 2D crystal phase and layer $B$ in 
a low density gas phase. To prepare the initial state for the dynamics, the layer separation is quenched to a small distance $s$, which leads to the formation of defects with various symmetries 
and patterns depending on the value of $s$ (see Fig.~\ref{fig:illustration}). (ii) We evolve the system to find glassy dynamics, identified by dynamical heterogeneity, the deviation from the linear 
diffusion law and a plateau in the relaxation (see Fig.~\ref{fig:resultspanel}). (iii) In
order to measure these features in a possible experiment with cold molecules we introduce  \textit{marker molecules}, i.e.~molecules prepared in a different internal state, which allow tracking the 
time evolution of the particles (see Fig. \ref{fig:marker}).

\textit{Preparation of the initial state} - We first prepare molecules in two uncoupled layers: a dipolar crystal of polar molecules in layer $A$ and a low density phase of defects in layer $B$. For 
uncoupled layers, the phase of layer $A$ is  described by parameters $r_d  = D m/\hbar^2 a$ and $K = k_B T a^3/D$, where $D=d_A^2/(4 \pi \epsilon_0)$. The first parameter is the ratio of the 
dipolar interaction $D/a^{3}$ for a given mean intralayer distance between the particles $a$, and the kinetic energy $\hbar^2/m a^{2}$, which in the dipolar crystal phase is $r_{d} \gg 1$. The second 
parameter measures the temperature in units of the interaction. For bosons the relevant phase diagram is sketched in Fig.~\ref{fig:illustration}(c) \cite{Buechler}, which at low temperatures shows 
a low density (2D) superfluid phase, a high-density crystal phase and a high temperature fluid phase. The theoretically predicted  conditions for the formation of a self-assembled crystal are, in the case of bosons, $r_d > 20$ and $K<0.1$ \cite{Buechler}. We note that these conditions for $r_{d}$ and $K$   are a requirement for temperature, density and dipole moment. While in present experiments with KRb the crystalline phase has not been realized, the ongoing effort in the laboratory to prepare cold ensembles of LiCs molecules \cite{LICS} with its large  electric dipole moment $\mu = 5.3 \text{Debye} $  and mass $m=92 \text{u}$ provides a promising candidate to obtain dipolar crystals.  In this case for a given temperature $T=0.1 \mu \text{K}$ the requirement on  $r_d$ and $K$ corresponds to an effective (induced) dipole moment  of  $d_A = 2.17 \text{Debye}$ and a lattice spacing of the self-assembled triangular crystal $a=0.32 \mu\text{m}$. We assume that the density in layer $B$ is lower than that in layer $A$ while all other parameters are the same. In the following we 
find it convenient to define length, energy and time in the reduced units $a$, $D/a^3$, $a \sqrt{m a^3/D}$, respectively.

To prepare the initial state for the glassy dynamics, the two layers are brought together by a quench in the layer separation. The speed of the quench is chosen such, that non-adiabatic effects are 
negligible \cite{supmat}. We consider two cases: the dipoles can be aligned parallel [see Fig. %
\ref{fig:illustration}(a)] or anti-parallel [see Fig. \ref{fig:illustration}%
(b)].  Fig. \ref{fig:illustration}(d) depicts the resulting defect patterns for parallel dipoles $d_B = 4 d_A$ where molecules belonging to different layers attract each other. This attraction
leads to stable dimers or trimers \cite{Baranov} if the two layers are separated by $s \leq
a$. These composite defects are reminiscent of interstitials in a crystal.
Here, the effective volume of the interstitials can be tuned by the ratio of
the dipole moments $d_A/d_B$ and the interlayer separation $s$.  For various layer separations ranging from $s=0.25$ to $s=a$
the resulting patterns include hexagonal, cubic and asymmetric defects. We note that the binding energy of  
similar impurities has been recently studied in a fermionic system \cite{Impurity}. 

In the anti-parallel case, $d_B = -d_A$, the interlayer interaction is
repulsive. This can be achieved using internal rotational degrees of freedom 
\cite{Baranov}. The defect patterns depicted in Fig. \ref{fig:illustration}%
(e) include cubic, five-fold symmetric and triangular symmetries reminiscent
of vacancy defects. Again, the effective volume of the defects can be tuned
by the layer separation. 

In both cases, parallel and anti-parallel bilayer setup, the combined system
is a mixture of effective defects and molecules. This is an experimental
realization of a binary mixture of dipoles which is a well known
glass-forming liquid in the classical regime \cite{KEIM2008,WEEKSGLASS}. We note, that
the anti-parallel setup allows one to induce defects with five-fold symmetry
which may allow one to implement a classical spin liquid model, where fivefold symmetries 
are studied \cite{FRUSTRATION}, in the quantum regime.

\textit{Glass Dynamics} - Below we study the relevant glass order parameters from the
dynamics of the proposed setup after the quench. In the classical regime we
use molecular dynamics simulations. In the quantum regime we employ the recently
developed dynamical path integral methods, which has been applied to the glass transition in Ref.~\cite{MARKLAND2011} (for
details on the numerical algorithms see SM \cite{supmat}). We note that these methods include quantum effects but ignore the exchange statistics between the particles and that we are interested mainly in the transition from the classical phase to a phase where quantum fluctuations become important. 

A  glass phase is identified by a dramatic increase in the relaxation time
which divergences with $\tau_{r} \propto \exp[-1/(T-T_g)]$ when approaching
the ideal glass temperature $T_g$ \cite{BINDERBOOK}. This corresponds to an
extended plateau in the self-intermediate scattering function \cite{BINDERBOOK}, a two-time correlation function defined as
\begin{equation}
F(k^*,\Delta t) = \frac{1}{N} \sum_j \langle e^{i k^* [%
\mathbf{r}_j(t) - \mathbf{r}_j(t + \Delta t) ] }\rangle.  \label{equ:fskt}
\end{equation}
Here, angular brackets $\langle \cdot \rangle$ denote ensemble averages over many realizations of the experiment after the quench. The sum runs over all particles in both layers $N = N_A + N_B$ and $k^* = |\mathbf{k}^*|$ is the absolute value of the characteristic $k$-vector of the coupled system
corresponding to the first peak in the static structure factor $S(\mathbf{k}) = \frac{1}{N} \sum_i \sum_j \langle e^{-i \mathbf{k} [\mathbf{r}_i - 
\mathbf{r}_j]} \rangle$ where $r$ is the distance in the $xy$-plane. Note
that Eq. (\ref{equ:fskt}) bears some similarity with the Fourier transform
of the density-density correlation function, however here, the sum runs over
individual particles at different times. In the path integral picture, the
analogous order parameter to the classical case Eq. (\ref{equ:fskt}) reads
as \cite{MARKLAND2011}

\begin{equation}
\Phi(k^*,\Delta t) = \frac{1}{N \hbar \beta} \int_{0}^{\hbar \beta} d\lambda
\langle \rho^\dagger(t+\Delta t+i\lambda) \rho(t)\rangle.
\end{equation}
Here, $\rho(\mathbf{k},\Delta t) = \sum_{j}^N e^{i \mathbf{k} \mathbf{r}_j}$ and the
integral runs over the imaginary path integral time. The semi-classical
approximation allows one to follow the real time propagation of the path
integral \cite{PARINELLO} neglecting exchange (for details see SM \cite{supmat}).

A crucial dynamical characteristic of the glass phase is its deviation from
the linear diffusion laws. In particular, the local diffusion of individual
particles is spatially heterogeneous, a characteristic of the glass phase
known as \textit{dynamical heterogeneity}. The local diffusion of
particle $j$ is $\Delta r_j(\Delta t) = \langle \mathbf{r}_j(t+\Delta t) - 
\mathbf{r}_j(t) \rangle$ and the mean squared displacements $\langle \Delta
r^2 (\Delta t) \rangle = 1/N \sum_i \Delta r_i^2(\Delta t)$. In the presence
of quantum fluctuations the analogous measure in the path integral picture
is the root mean squared displacement of the projections of the path
integral with $\hat{r}_j(t) = \frac{1}{N \hbar \beta} \int_{0}^{\hbar \beta}
d\lambda r_j(t+i\lambda)$ and  $\Delta \hat{r}_j(\Delta t) = \langle \mathbf{\hat{r}}_j(t+\Delta t) - 
\mathbf{\hat{r}}_j(t) \rangle$. 

The unique feature of the bilayer system of ultracold molecules is the possibility to study glassy dynamics in both the classical and quantum regime, which are characterized by the dominance of thermal vs.~quantum fluctuations, respectively. This transition can be controlled  by three tunable parameters: $s$ representing the layer separation, the dimensionless temperature $K$, and the dipolar crystal parameter $r_d$. The case of large $r_{d} \equiv D m/\hbar^2 a$ corresponds to the classical limit [compare  Fig.~\ref{fig:illustration}(c)]. The associated classical glass transition can be studied by varying $K$ with fixed $r_d$, while varying the initial defects with $s$ [see  Fig.~\ref{fig:illustration}(d) and (e)]. Below we will discuss this limit in Fig.~\ref{fig:resultspanel}(a,b). The transition from the classical to the quantum regime is achieved by decreasing $r_{d}$. This increasing role of quantum fluctuations on the glassy dynamics is summarized in Fig.~\ref{fig:resultspanel}(c) for various $r_{d}$, while Fig.~\ref{fig:resultspanel}(d) discusses the effect to defect patterns, as controlled by $s$ [compare  Fig.~\ref{fig:illustration}(d) and (e)]. 

The averaged root mean square displacement as a function of time is shown in Fig.~\ref{fig:resultspanel}(a). The dynamics does not
follow the Einstein diffusion law $\langle \Delta r^2(\Delta t) \rangle \neq
2 D d \Delta t$, where $d$ is the dimension of the system, but develops a plateau larger than the sampling time, which corresponds to a divergence in the thermodynamic limit, for an effective $K$ of
the combined system smaller than $K<0.02$. Fig.~\ref{fig:resultspanel}(b)
shows the according relaxation dynamics $F(k^*,\Delta t)$ for a final layer
separation $s=0.5a$ in the classical regime. Both setups with parallel and
anti-parallel dipoles show a growing plateau and enter a glass phase.

Including quantum fluctuations the relaxation dynamics of the system
changes considerably. Fig.~\ref{fig:resultspanel}(c) depicts the relaxation
dynamics $\Phi(k^*,\Delta t)$ calculated from a path integral simulation as
a function of $r_d$ at a fixed $K$. With increasing $r_d$, the
dynamics approaches that of the classical molecular dynamics simulation
corresponding to $r_d \rightarrow \infty$ where the system is glassy. By lowering $r_d$,
the system melts due to large quantum fluctuations. In the
bilayer system, one can also tune the defect size with the layer separation
(see Fig.~\ref{fig:illustration}). Fig.~\ref{fig:resultspanel}(d) shows a
set of parameters, where one can reach the glass phase by variation of the
layer separation. In contrast to the classical model system of colloids \cite{KEIM2008}, the actual
dynamics after the initial thermalization corresponds to that of an isolated
system (micro-canonical). However, this does not influence the long tail
part in the relaxation which is independent of the microscopic dynamics \cite{GLEIM}. 

\textit{Measurement of Dynamical Heterogenity} - One way to measure the anomalous diffusion as a glass feature is 
to consider a  subensemble of polar molecules in a small spatial region ({\em marker molecules}), which are addressed with a laser and transferred to another internal (e.g.~hyperfine) state  \cite{OSPELKAUS}. The position of this molecular cloud at a later time can be measured with optical techniques. The ultimate tool for such position measurements will be provided by the quantum gas microscope for molecules, which - following the atomic case - will provide single site / single molecule detection with a resolution of the single site of an optical lattice \cite{PRIVATECOMM}. 
Dynamical heterogeneity can then be measured as follows: We mark molecules in the crystal layer A in a small region with laser waist-size $R_0 > a$. 
After the system  evolves in time, the positions of the {\em marker molecules} is measured, and
the extension of the cloud of marker molecules is given by the radius of gyration $R_g^2=1/(2N^2)
\sum_{i,j} ({\mathbf{r}}_i-{\mathbf{r}}_j)^2$. In a liquid a cloud of tagged particles
spreads out linearly with time while in an amorphous
solid it is constant $R_g/R_0=1$. Most important, in a liquid and crystal, the
dynamics is independent of the initial position of the makers molecules as the
dynamics is spatially homogeneous.
However, in the glass phase due to dynamical heterogeneity the scaling differs significantly depending on
the initial position (see Fig.~\ref{fig:marker}). As the initial size is $R_0$ from the laser waist, a series of single (destructive) measurements after time $\Delta t$ allows one to distinguish a glass from liquid or amorphous solid  (see Fig.~\ref{fig:marker}).  Note that the marker
molecules are quantum mechanically distinguishable which changes the dynamics in the deep quantum regime when exchange statistics is included, which is not relevant in the present case. 

In conclusion, we have shown that a bilayer system of polar molecules can,
with properly chosen parameters, exhibit a glass phase in a regime where quantum fluctuations are of the order of thermal fluctuations. Thus ultracold ensembles of polar molecules could provide a widely tunable paradigmic model system for classical and quantum glass physics, where theory and experiment can meet in a yet unexplored parameter regime of glass physics, providing in particular a stimulus for the theoretical developments.

We thank K.~Binder and M.~Baranov for fruitful discussions. Work supported by the Austrian Science Fund (FWF): P 25454-N27, SFB FOQUS, Marie Curie Initial Training Network COHERENCE,  and
 ERC Synergy Grant UQUAM.

\newpage

\newpage
\onecolumngrid

{
\center \bf \Large 
Supplemental Material to\vspace*{0.1cm}\\ 
\emph{From Classical To Quantum Glasses with Ultracold Polar Molecules
}
}

\section{Numerical Methods}

\subsection{Classical Dynamics}

We use a second-order Langevin integrator \cite{EIJNDEN} to study the dynamics of the bilayer system in the classical regime. The time evolution is given by the Langevin equation 
\begin{equation}
\label{equ:n}
m \frac{\partial^2 \mathbf{x}}{\partial t^2} = - \nabla V(\mathbf{x}_1,...,\mathbf{x}_1) - \gamma m \frac{\partial \mathbf{x}}{\partial t} + \sqrt{2 k_B T \gamma m} \eta(t).
\end{equation}
Here, $W$ corresponds to a Wiener process and $\eta(t) = \dot{W}$. This equation can be solved numerically in second order accuracy with the following updates

\begin{eqnarray}
\label{equ:eulermarayuma}
v(t + \Delta t/2) &=& v(t) + \frac{1}{2} dt F(\mathbf{x}(t)) - \frac{1}{2} dt \gamma v(t) + \\ \nonumber 
&+&\frac{1}{2} \sqrt{dt} \sigma \zeta^n - \frac{1}{8} dt^2 \gamma (F(\mathbf{x}(t)) - \gamma v(t)) - \\ \nonumber 
&-&\frac{1}{4} h^{3/2} \gamma \sigma (\frac{1}{2} \zeta^n + \frac{1}{\sqrt{3}} \eta^n ) \\ 
x(t + \Delta t) &=& x(t) + dt v(t + \Delta t/2) + dt^{3/2} \sigma \frac{1}{2 \sqrt{3}} \eta^n \\ 
v(t + \Delta t) &=& v(t + \Delta t/2) + \frac{1}{2} dt F(\mathbf{x}(t + \Delta t)) - \\ \nonumber 
&-&\frac{1}{2} dt \gamma v(t + \Delta t/2) + \frac{1}{2} \sqrt{dt} \sigma \zeta^n - \\ \nonumber  
&-& \frac{1}{8} dt^2 \gamma (F(\mathbf{x}(t+\Delta t)) - \gamma v(t + \Delta t/2)) - \\ \nonumber 
&-&\frac{1}{4} h^{3/2} \gamma \sigma (\frac{1}{2} \zeta^n + \frac{1}{\sqrt{3}} \eta^n ) \\ 
\end{eqnarray}
Here, $\eta$ and $\zeta$ are two independent noise variables from a Wiener process and $F(\mathbf{x}(t) = \sum_{ij} \nabla V_{ij}$. Note that in the limit of zero-friction $\gamma \rightarrow 0$, the equations of motion reduce to a energy conserving (micro-canonical) propagator. 

\subsection{Semi-classical Dynamics}

For completeness we repeat here the derivation of the numerical integrator used in the semi-classical regime given in Ref. \cite{PARINELLO}. The partition function is rewritten in terms of Feynman path integrals
\begin{equation}
\langle A \rangle = \frac{1}{Z}tr[e^{-\beta H} A],
\end{equation}
with
\begin{equation}
Z = \frac{1}{(2 \pi \hbar)^f} \int d^f \mathbf{p} \int d^f \mathbf{q} e^{-\beta_n H_n(\mathbf{q},\mathbf{p})}.
\end{equation}
Here, $f=Nn$ and $\beta_n = \beta/n$. The full Hamiltonian consists of system Hamiltonian and interaction Hamiltonian

\begin{equation}
H_n(\mathbf{q},\mathbf{p}) = H_n^0(\mathbf{q},\mathbf{p}) + V_n(\mathbf{q}).
\end{equation}
The system part is 
\begin{equation}
H_n^0(\mathbf{q},\mathbf{p}) = \sum_{i=1}^{N}\sum_{j=1}^n \left(\frac{(p_i^{(j)})^2}{2 m_i} + \frac{1}{2}m_i \frac{1}{\beta_n^2 \hbar^2} [q_i^{(j)} - q_i^{j-1}]^2 \right).
\end{equation}
While sophisticated semi-classical methods have been developed to sample this Hamiltonian without dissipation, the coupling to a thermal bath is challenging as the frequencies in the path space separate from the frequencies from the thermal bath. In the framework of path integral Langevin dynamics, this is solved by splitting the dynamics into two parts: First, the full Hamiltonian evolves according to the path integral formalism in absence of a thermal bath. In a second step, the thermal bath acts on the slowest and higher normal modes of the paths separately, on the level of the equations of motion. 

The energy conserving Liouvillian part of the dynamics (without dissipation) is written as a sum of system and interaction 

\begin{equation}
L = L_0 + L_V.
\end{equation}
Here, $L_0 = -[H_n^0(\mathbf{q},\mathbf{p}),.]$ and $L_V = -[V_n^0(\mathbf{q},\mathbf{p}),.]$. Using Totter's theorem, the propagator for a small $\Delta t$ can be written as

\begin{equation}
e^{\Delta t L} = e^{(\Delta t/2)L_V}e^{\Delta t L_0}e^{(\Delta t/2)L_V}.
\end{equation}
The semi-classical treatment of the heat bath requires the normal mode represent of the Hamiltonian which is derived from the transformation

\begin{eqnarray}
\tilde{p}_k = \sum_{j=1}^n p_i^{(j)} C_{jk} \;  {\text and} \;\tilde{q}_k = \sum_{j=1}^n q_i^{(j)} C_{jk},
\end{eqnarray}
where the matrix elements of $c_{jk}$ are

\begin{equation}
\label{wolf}
  C_{jk}=\begin{cases}
    \sqrt{1/n}, & \text{if $k=0$}\\
    \sqrt{2/n} \cos(2 \pi jk/n), & \text{if $1 \leq k \leq n/2-1$}\\
    \sqrt{1/n}(-1)^j, & \text{if $k = n/2$}\\
    \sqrt{2/n} \sin(2 \pi jk/n), & \text{if $n/2+1\leq k\leq n -1$}.
  \end{cases}
\end{equation}
In the normal mode representation the Hamiltonian reads as

\begin{equation}
H_n^0(\mathbf{q},\mathbf{p}) = \sum_{i=1}^{N}\sum_{k=0}^{n-1} \left(\frac{(\tilde{p}_i^{(j)})^2}{2 m_i} + \frac{1}{2}m_i \frac{\sin(k \pi/n)}{\beta_n^2 \hbar^2} (\tilde{q}_i^{k})^2 \right).
\end{equation}
Dissipation is then added on the level of the equations of motion in the space of normal modes. This leads to the following equations of motion 

\begin{equation}
e^{\Delta t L} = e^{(\Delta t/2)L_\gamma}e^{(\Delta t/2)L_V}e^{\Delta t L_0}e^{(\Delta t/2)L_V}e^{(\Delta t/2)L_\gamma}
\end{equation}
While $L_0$ and $L_V$ are associated with momentum and position dynamics in the path integral, $L_\gamma$ is defined in the space of normal mode variables. This requires four transformation between normal mode and path integral space in each time step. The Trotter step associated with $L_V$ is 

\begin{equation}
p_i^{(j)}(t + \Delta t) = p_i^{(j)}(t) - \Delta t \frac{\partial V}{\partial q_i^{(j)}}
\end{equation}
The evolution of the path integral with $L_0$ defined in normal mode space corresponds to 

\begin{equation}
\tilde{p}_i^{(k)}(t + \Delta t) = \\
\tilde{p}_i^{(j)}(t) \cos(\omega_k \Delta t) - \tilde{q}_i^{(j)}(t) m_i \omega_k \sin(\omega_k \Delta t).
\label{eq:1}
\end{equation}
and
\begin{equation}
\tilde{q}_i^{(k)}(t + \Delta t) = \\
\tilde{p}_i^{(j)}(t)\frac{1}{m_i \omega_k} \sin(\omega_k \Delta t) - \tilde{q}_i^{(j)}(t) \cos(\omega_k \Delta t).
\label{eq:2}
\end{equation}
In the last step, $L_\gamma$ describes thermalization in the normal mode space and leads to the updates 

\begin{equation}
\tilde{p}_i^{(k)}(t + \Delta t) = \tilde{p}_i^{(k)}(t) e^{-\Delta t/2 \gamma^{(k)}} + \sqrt{\frac{m_i}{\beta_n} (1 - e^{-\Delta t \gamma^{(k)}})} \xi_i^{(k)}.
\end{equation}
\begin{figure}[ht]
\centerline{\includegraphics[width=8.5cm]{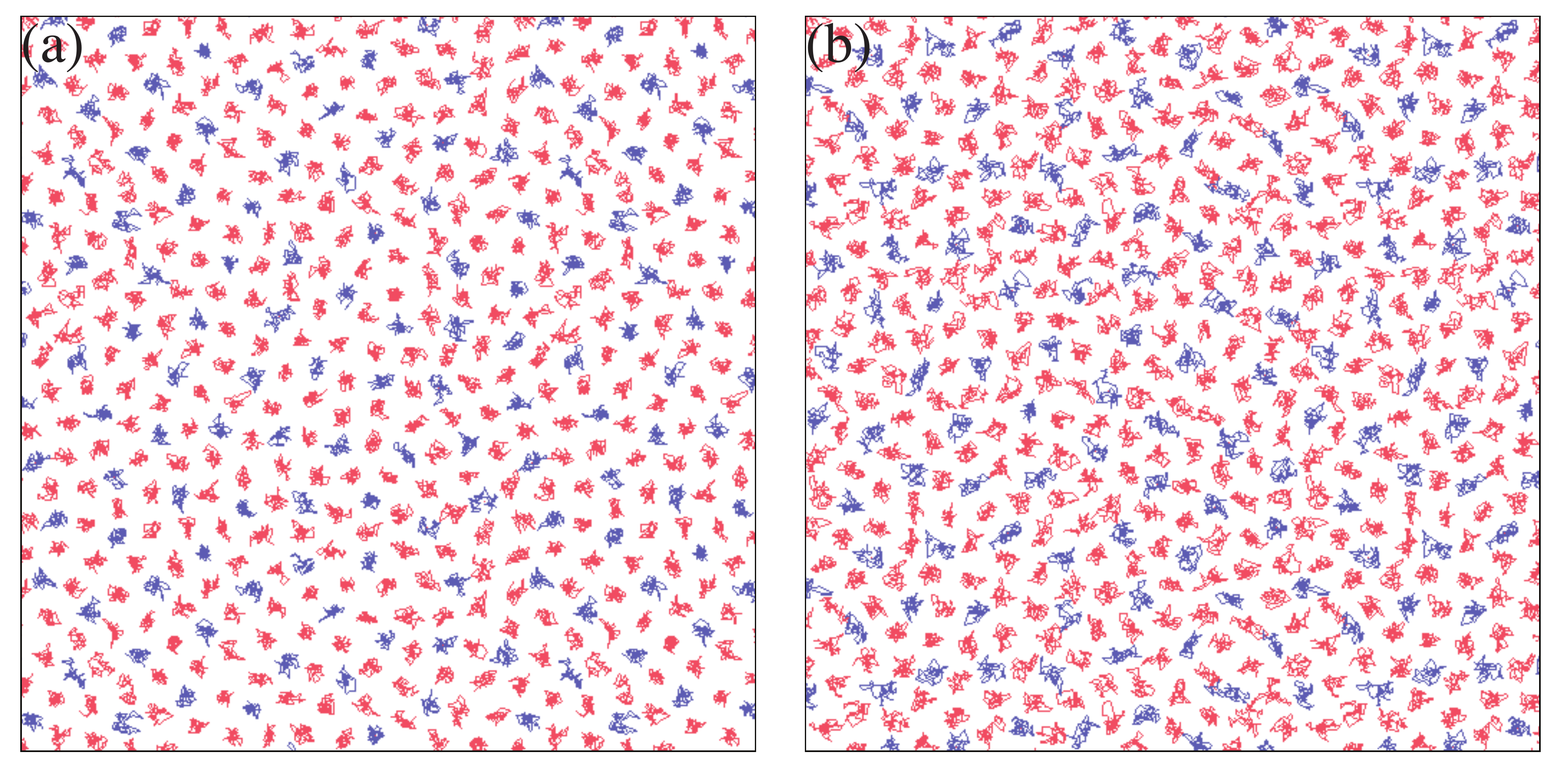}}
\caption{Snapshots from a path integral simulation with $d_B=-d_A$, $K=0.02$ deep in the crystalline region $r_d=40$ (a) and close to the superfluid region $r_d=25$ (b) (see Fig. 2 in the main text) . The projection of the paths onto real space of layer $A$ (red) and layer $B$ (blue) indicate the extension of the wave function.}
\label{fig:paths}
\end{figure}
Fig. \ref{fig:paths}a and b show typical snapshots from a path integral simulations in the large $r_d$ limit (classical) compared to a phase where quantum fluctuations become important. 

\section{Structure Factor and Intermediate Scatter function}

The dynamical slowing down of the glass phase corresponds to the development of a plateau in the time dependent intermediate scatter function $F(k,t)$ (classical) and $\Phi(k,t)$ (path integral). This time dependent order parameter allows one to measure the relaxation time of a characteristic $k$-vector. This characteristic $k$-vector ($k^*$) is associated with the first peak in the static structure factor $S(k)$ and taken as an input parameter for the time dependent intermediate scatter function of the Kubo transformed density $\Phi(k^*,t)$ (or $F(k^*,t)$).
The static structure factor $S(\mathbf{k})$ is defined as

\begin{equation}
S(\mathbf{k}) = \frac{1}{N} \langle \sum_{l,m} e^{-i{\bf \mathbf{k}} ({\bf R}_l - {\bf R}_m)} \rangle.
\end{equation}
Here, $R_i$ is the position of the classical particle or in the quantum case the centroid of the path integral $R_i = 1/n \sum_n r_i^n$, where the average runs over all imaginary time slices. Finally, the time-dependent density-density correlation function is 

\begin{equation}
\Phi(k^*,\Delta t) = \frac{1}{N \hbar \beta} \int_{0}^{\hbar \beta} d\lambda \langle \rho^\dagger(t+\Delta t+i\lambda) \rho(t)\rangle
\end{equation}
Here, $\rho(k,t) = \sum_{j}^N e^{i \mathbf{k} \mathbf{r}_j}$ and the integral runs over the imaginary path integral time. The ensemble averages $\langle \cdot \rangle$ are taken over a large number of realizations. 

\begin{figure}[ht]
\centerline{\includegraphics[width=8cm]{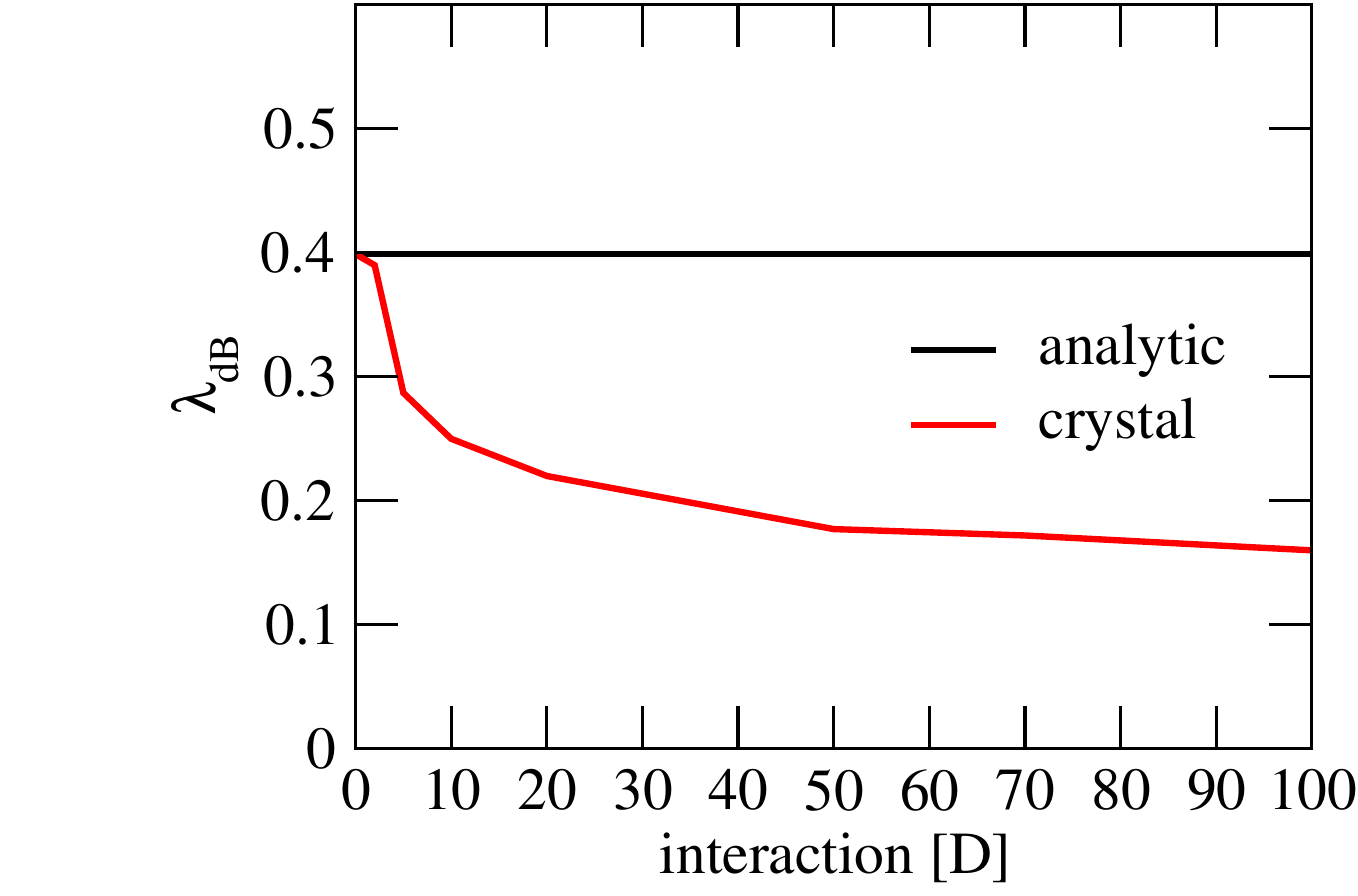}}
\caption{The extension of the particle wave function of a free particles with unit temperature and mass is in units of $\hbar$: $\lambda_{dB} = \sqrt{1/2 \pi} = 0.3989$ (red). In the crystal with average particle distance $a=1$ (density $\rho = 2/\sqrt{3}$), the pair interactions localize the particles resulting in a decrease in the wavefunction extension.}
\label{fig:wavelength}
\end{figure}

\section{Comparison of Path integral and Wigner function formalism}

The extension of the wavefunction of a free particle with mass $m$ at temperature $T$ is 
\begin{equation}
\lambda_{dB} = \sqrt{\frac{h^2}{2 \pi m k_B T}}. 
\end{equation}
In the path integral formalism, the average extension of the particles is calculated from
\begin{equation}
\lambda = 2 \langle \frac{1}{N \hbar \beta} \int_0^{\hbar\beta} d\lambda \left| \mathbf{r}(t+i \lambda) - \hat{\mathbf{r}}_j(t) \right|^2 \rangle.
\end{equation}
To check for consistency we compare $\lambda_{dB}$ with the numerical results. Fig. \ref{fig:wavelength} depicts the particle extension as a function of the interaction strength $D$. In the free particle limit $D \rightarrow 0$ the result approaches exactly the expected value of $\lambda_{dB}$. The extension of the wavefunction can also be calculated from the width of the Wigner function with a thermal characteristic function after integrating out the momentum. The Wigner function of a thermal state is \cite{QN}

\begin{eqnarray}
\label{equ:wigner}
W(\alpha,\alpha^*) &=& \\   \nonumber
&=& \frac{1}{\pi^2} \int d^2 \lambda \exp[-\lambda \alpha^* + \lambda^* \alpha] \times \\ \nonumber
& &\exp[-\frac{1}{2} |\lambda|^2] \exp[\frac{-|\lambda|^2}{e^{\hbar \omega/ k_B T}-1}] \\  \nonumber
&=& \frac{2}{\pi} \tanh[\frac{\hbar \omega}{2 k_B T}] \exp\{-2 |\alpha|^2  \tanh[\frac{\hbar \omega}{2 k_B T}]\}
\end{eqnarray}
Using 
\begin{equation}
\alpha = \frac{i p}{\hbar} \frac{1}{\sqrt{2 \eta}} + x \sqrt{\frac{\eta}{2}}
\end{equation}
with 
\begin{equation}
\eta = \sqrt{k m}/\hbar
\label{equ:eta}
\end{equation}
we rewrite Eq. \ref{equ:wigner} in terms of position and momentum $W(x,p) = \frac{1}{2 \hbar} W(\alpha,\alpha^*)$. Integrating out the momentum leads to

\begin{eqnarray}
\label{equ:wignerx}
W(x) &=& \int_{-\infty}^{\infty} W(x,p) dp = \\ \nonumber
&=& \sqrt{\frac{\eta}{2 \pi}} \sqrt{\tanh[\frac{\hbar \omega}{k_B T}]} \exp\{-\tanh[\frac{\hbar \omega}{k_B T}] \eta x^2\}
\end{eqnarray}
We interpret the Wigner function as a quasi-probability and compare $W(x)$ to the Gaussian distribution of a thermal wavefunction 

\begin{equation}
P(x) = \frac{1}{\sqrt{2 \pi \lambda_{dB}^2}} \exp[-\frac{x^2}{2 \lambda_{dB}^2}]
\label{equ:px}
\end{equation}
Expanding the $\tanh$ in Eq. (\ref{equ:wignerx}) and comparing the coefficients of Eq. \ref{equ:wignerx} and \ref{equ:px} we find

\begin{equation}
\frac{\hbar^2}{2 \pi m k_B T} = \frac{\hbar \omega}{k_B T} \eta
\label{equ:px}
\end{equation}
which is consistent with Eq. (\ref{equ:eta}).

In a crystal the potentials of neighboring sites act as a trapping potential which localizes the wavefunctions (see Fig. \ref{fig:wavelength}). 

\begin{figure*}[ht]
\centerline{\includegraphics[width=14cm]{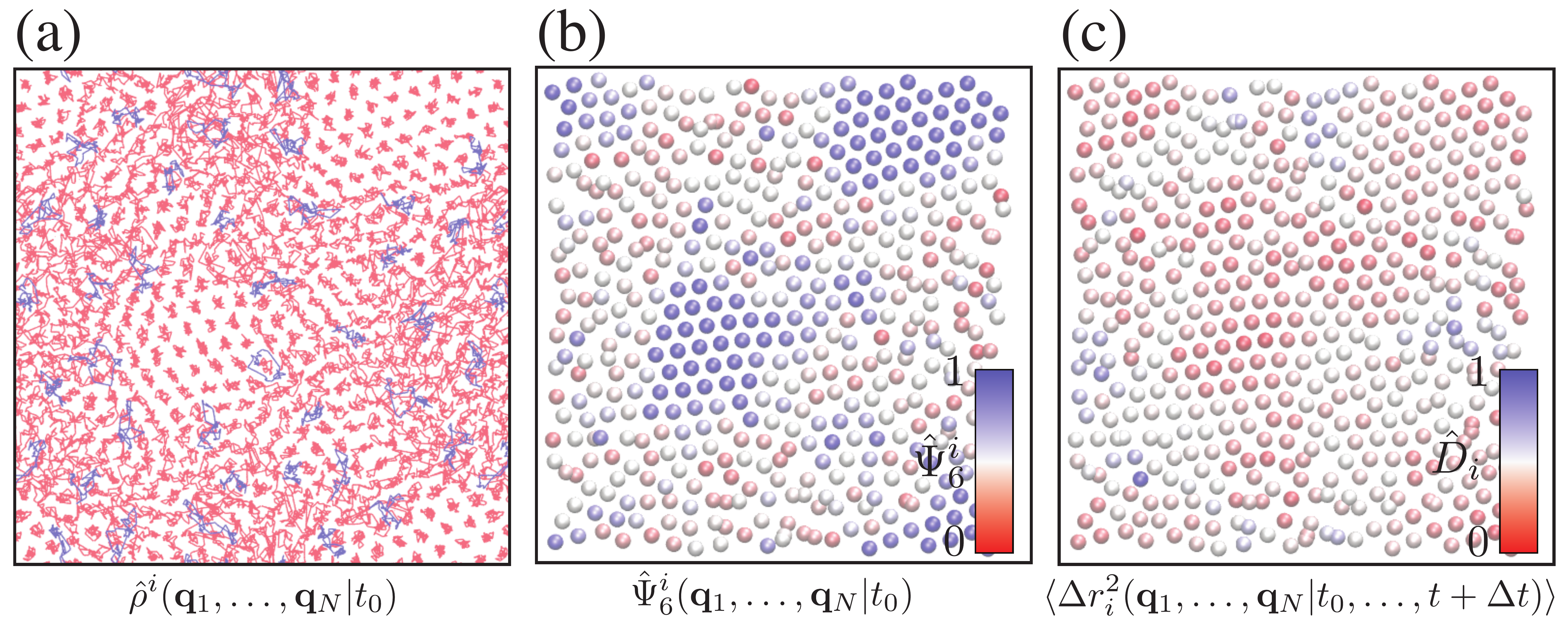}}
\caption{(Color online) Three different representations of a typical configuration in the glass phase with partial crystalline order. (a) Projection of the path integral density to real space with particles in the crystal layer (red) and particles from the defect layer (blue). The regions of large delocalization are correlated with defect particles while particles in crystalline regions are more localized. (b) Maximum of the probability positions of particles as spheres colored by the value of the time independent medium range order parameter $\hat{\Psi}_6$ ranging from unordered regions (red) to maximal ordered regions (blue). (c) Particles colored by the time dependent mean squared displacement  $\hat{D}_i = \frac{D_i - D_{\textrm{min}}}{D_{\textrm{max}} - D_{\textrm{min}}}$ ranging from frozen regions (red) to regions of large displacement (blue) .}
\label{fig:heterogeneity}
\end{figure*}

\section{Heating due to the quench}

The system is described by the Hamiltonian Eq. (1) in the main text. Here, we explicitly include the time dependent parameter $\lambda$ which allows one to vary the interlayer separation

\begin{figure}[ht]
\centerline{\includegraphics[width=8cm]{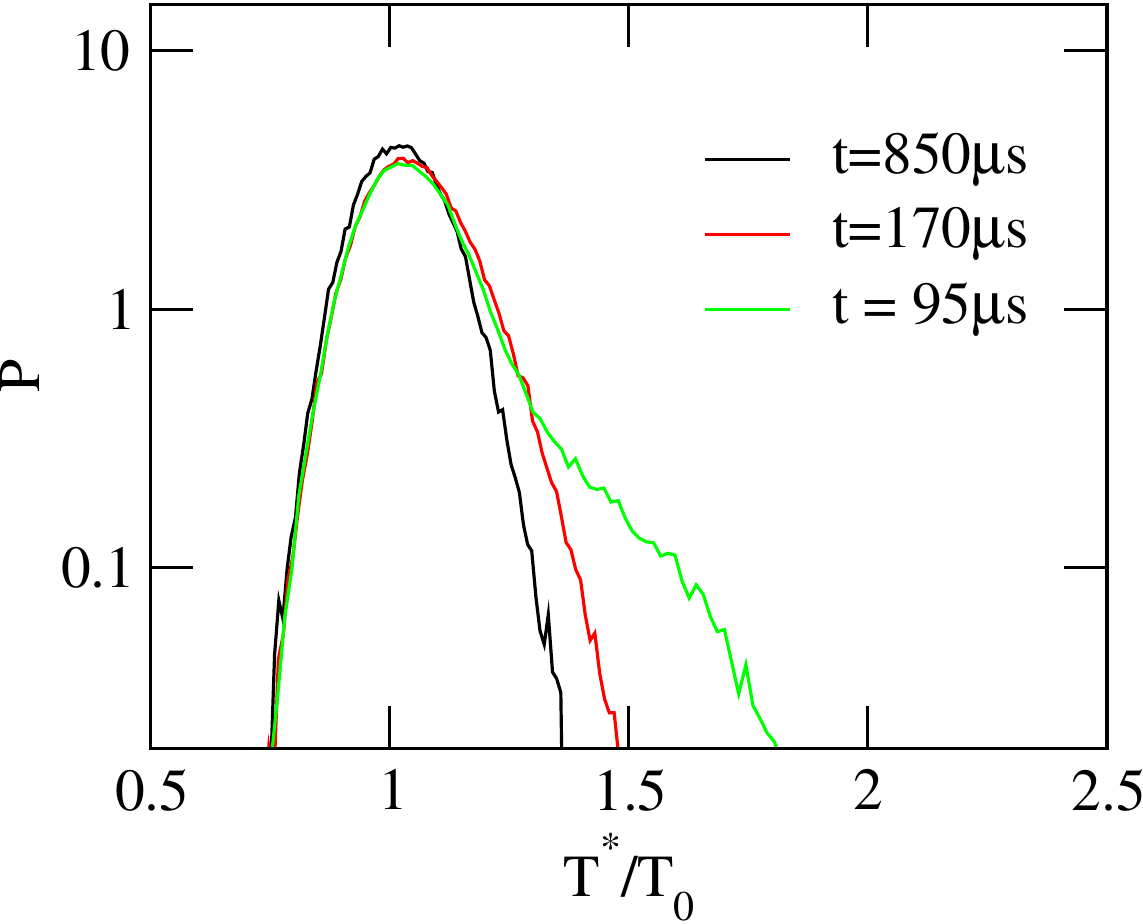}}
\caption{Probability distribution of the effective temperature $T^*/T_0$ after a quench from $s(0)=a$ to $s(t)=0.5a$ for various quench velocities. Densities and temperature a chosen as in case anti-parallel dipoles depicted in Fig. 1 in the main text. For total quench times larger than $t=850\mu s$ the temperature is basically unchanged while a fast quench with $t<170\mu s$ leads to considerable non-adiabatic dissipative work. The distributions is sampled from $50000$ runs.}
\label{fig:heating}
\end{figure}

\begin{eqnarray}
\label{equ:hamiltonian}
H(\lambda(t)) = \sum_i \frac{p^2}{2m} + \sum_{A} \frac{D}{\sqrt{x^2 + y ^2}^3} + \\ \nonumber
\sum_{B} \frac{D}{\sqrt{x^2 + y ^2}^3} + \sum_{i,j} \frac{D \frac{3s(\lambda(t))^2}{x^2 + y^2 + s(\lambda(t))^2} - 1 }{\sqrt{x^2+y^2+s(\lambda(t)^2}^3}.
\end{eqnarray}
Here, the sum over $A$ and $B$ indicates interactions between particles of the same layer and the last term the interaction between particles in different layers. The layer separation is then linearly switched from $s_{\text init}$ to $s_{\text final}$ with

\begin{equation}
s(\lambda(t)) = \lambda(t) s_{\text init} + (1-\lambda(t))s_{\text final}
\end{equation}
by the protocol 

\begin{equation}
\lambda(t) = v_l t.
\label{equ:protocol}
\end{equation}
The non-equilibrium work performed during the quench is the sum of reversible work and irreversible work $W = W_{\textrm rev} + W_{\textrm irr}$ or explicitly 
\begin{equation}
W = \int_0^1 d\lambda \frac{\partial \lambda}{\partial t} \frac{\partial H(\lambda(t))}{\lambda} 
\end{equation}
The irreversible work from the quench increases with the speed of the protocol and heats up the system non-adiabatically. The protocol is a follows: First, the molecules are prepared at temperature $T_0$ as described in the main text. By a rapid quench in the interlayer separation, the system is driven out of equilibrium. During the quench, the system evolves micro-canonically and work is performed from the change in the layer separation $s(\lambda)$. We define an effective temperature in the micro-canconical ensemble as the average kinetic energy 
\begin{equation}
T^* = \frac{2 K }{N d k_B}.
\end{equation}
Here, $K = \frac{N m \langle v^2 \rangle}{2}$ is the kinetic energy and $d=2$ the dimensionality of the system. Fig. \ref{fig:heating} depicts the probability density of the effective temperature in the final state. For a slow quench with switching time larger than $t>170\mu s$ the temperature increase is negligible. 

\section{Dynamical Heterogeneity vs. Structural order}

The local diffusion $D_i = \langle \left[ \mathbf{r}_i(t + \Delta t) - \mathbf{r}_i(t)\right]^2 \rangle$ reveals a remarkable connection between structure and dynamics in the glass phase as was recently shown in colloidal glasses \cite{TANAKACRITICAL,TANAKACRITICAL2}. In particular, for some parameters the mobility of individual particles $D_i$ is correlated with the local structure of the neighbors. This local order is measured in two dimensions by the averaged  hexatic order parameter 
\begin{equation}
\hat{\Psi}_6=\frac{1}{N_b} \sum_{N_b} |\Psi_6^i|^2 
\end{equation} 
with 
\begin{equation}
\Psi_6^i = \frac{1}{N_b} \sum_{N_b} e^{i 6 \phi_{ij}}.
\end{equation}
Here, the sum runs over all neighbors $N_b$ including all particles within a distance $r_n$ of particle $i$. 

The dynamical $D_i$ and the structural order parameter $\hat{\Psi}_6^i$ are homogeneous in a liquid or a solid but are locally heterogeneous and anti-correlated in the glass phase. Fig.~\ref{fig:heterogeneity}(b) and (c) show the mean position of the particle wavefunctions where colors represent the value of $\hat{\Psi}_6^i$ and $\hat{D}_i$, respectively, with the normalized diffusion $\hat{D}_i = \frac{D_i - D_{\textrm{min}}}{D_{\textrm{max}} - D_{\textrm{min}}}$. 

In the path integral picture, structure and dynamics is also correlated with the extension of the particles in imaginary time. Fig.~\ref{fig:heterogeneity}(a) shows the projection of the paths from the path integral simulation. The extension of the quantum fluctuations are  calculated from the second moment of the normal modes of the path integral $\lambda =  2 \langle \frac{1}{N \hbar \beta} \int_0^{\hbar\beta} d\lambda \left| \mathbf{r}(t+i \lambda) - \hat{\mathbf{r}}_j(t) \right|^2 \rangle$ which is consistent with the analytic result of the extension of the Wigner function from a thermal quantum characteristic function (see section III). The extension of a free particle at finite temperature is given by the thermal deBroglie wavelength $\lambda = [h^2/(2 \pi k_B T m)]^{1/2}$ while in a crystal the extension is smaller due to the interaction of neighboring particles acting as a trapping potential which is to first order quadratic. In the glass, the structure and therefore the effective potential at each site is heterogeneous which is also apparent in the wavefunction extension. 

\bibliographystyle{prsty}

\end{document}